\documentclass[runningheads]{llncs}
\usepackage[T1]{fontenc}
\usepackage{graphicx}

\usepackage[hyphens]{url}
\usepackage{hyperref}
\usepackage[hyphenbreaks]{breakurl}
\usepackage{color}

\usepackage[inline]{enumitem}
\usepackage{array}
\usepackage{booktabs}
\usepackage{xcolor}
\usepackage{afterpage}
\usepackage{pdflscape}

\begin{document}
\title{The Viability of Continuous Experimentation in Early-Stage Software Startups}
\subtitle{A Descriptive Multiple-Case Study}
\titlerunning{The Viability of CE in Early-Stage Software Startups}

\author{Vihtori M{\"a}ntyl{\"a}\orcidID{0000-0002-0448-9494}
\and
Bettina Lehtelä\orcidID{0000-0002-2814-4386}
\and
Fabian Fagerholm\orcidID{0000-0002-7298-3021}}
\authorrunning{V. M{\"a}ntyl{\"a} et al.}
%
\institute{Aalto University, P.O.\ Box 15400, Espoo, Finland \\
\email{vihtori.mantyla@protonmail.com},
\email{bettina.lehtela@aalto.fi},
\email{fabian.fagerholm@aalto.fi}}

\maketitle              

\begin{abstract}
\textbf{Background:}
Continuous experimentation (CE) has been proposed as a data-driven approach to software product development. Several challenges with this approach have been described in large organisations, but its application in smaller companies with early-stage products remains largely unexplored.
\textbf{Aims:}
The goal of this study is to understand what factors could affect the adoption of CE in early-stage software startups.
\textbf{Method:}
We present a descriptive multiple-case study of five startups in Finland which differ in their utilisation of experimentation.
\textbf{Results:}
We find that practices often mentioned as prerequisites for CE, such as iterative development and continuous integration and delivery, were used in the case companies. CE was not widely recognised or used as described in the literature. Only one company performed experiments and used experimental data systematically.
\textbf{Conclusions:}
Our study indicates that small companies may be unlikely to adopt CE unless
\begin{enumerate*}[label=\arabic*)]
\item at least some company employees have prior experience with the practice,
\item the company's limited available resources are not exceeded by its adoption, and
\item the practice solves a problem currently experienced by the company, or the company perceives almost immediate benefit of adopting it.
\end{enumerate*}
We discuss implications for advancing CE in early-stage startups and outline directions for future research on the approach.

\keywords{continuous experimentation, new product development, startup, continuous software engineering, case study}
\end{abstract}


\section{Introduction}

A mismatch between product features and customer needs is one of the most common reasons for software startup failure~\cite{klotins2019}. While agile methods emphasise customer value~\cite{dingsoyr2016}, building the right product appears to be a feat that few startups achieve. Continuous experimentation (CE) is a software engineering method where product development is driven by field experiments with real users~\cite{holmstromolsson2012,fabijan2017b,fagerholm2017,yaman2017,auer2021}. It strives to establish virtuous feedback loops between business, development, and operations~\cite{fitzgerald2017}, and reportedly improves product quality and business performance~\cite{fabijan2017a,fabijan2017b}, with promising implications for startups.

Previous studies have found that startups tend to run experiments as part of their product development process ~\cite{lindgren2015,lindgren2016}. 
However, the experiments are usually neither planned nor run in the organised and systematic manner characteristic of CE~\cite{lindgren2015,melegati2019}.
Startups are often reluctant to incorporate industry best practices, methods, and frameworks proposed by researchers into their business and product development processes~\cite{klotins2019}. This raises the question of whether CE, as envisioned in different frameworks and models in the literature, fits the needs of small companies with early-stage products, such as software startups.

This paper aims to elucidate possible reasons for why adopting CE in early-stage startup (ESS) settings might be difficult. 
Rather than producing a general theory of adoption of CE in startups, we develop propositions that can be extended and validated in future studies, and that can point to novel directions for research. 
Specifically, we address the following research questions:

\begin{enumerate}[label=\textbf{RQ\arabic*: },left=1em]
\item Do ESS companies integrate experimentation into their software development practices?
\item How do ESS companies utilise user data in their software product development process?
\item What factors influence the adoption of CE in ESS companies?
\end{enumerate}

We study these questions in the context of five startup companies. We contribute
\begin{enumerate*}[label=\arabic*)]
\item a description of the state of experimentation and overall mode of operation in the case companies from the perspective of CE, providing insight into what product development practices startups tend to adopt and why; and
\item a set of factors that affect the adoption of CE in startup companies, partly supporting earlier findings from startups (e.g.,~\cite{lindgren2015,lindgren2016,gutbrod2017,melegati2019,melegati2022}) and larger companies (e.g.,~\cite{yaman2017}), but with new and more detailed factors relevant to adoption of CE.
\end{enumerate*}
The paper builds on results obtained in a Master's thesis~\cite{mantyla2022}.

\section{Background}

CE approaches software product development through experiments with real users~\cite{holmstromolsson2012,fabijan2017b,fagerholm2017,yaman2017,auer2021}.
This includes collecting and analysing experimental data to test product hypotheses, gaining insights for new feature ideas to be evaluated in subsequent experiments.
CE comes with various implications on organisational structure and culture~\cite{kohavi2009,fitzgerald2017,yaman2017}, product and software development processes~\cite{holmstromolsson2012,fagerholm2017}, and software architecture, construction, and deployment~\cite{fagerholm2017,schermann2018}.

\subsection{Scientific and Online Controlled Experiments}

Scientific experiments can provide causal descriptions of what consequences varying a treatment has on a subject of interest~\cite{shadish2001}. Experimental setups vary with the study they are used for, but they all share the same basic elements:
\begin{enumerate*}[label=\arabic*)]
\item a hypothesis to be tested, including evaluation criteria for its acceptance or rejection; 
\item sufficient control of independent variables; 
\item means for collecting data; and 
\item experimental subject(s) under investigation~\cite{shadish2001}.
\end{enumerate*}

Online controlled experiments (OCE)~\cite{kohavi2013} leverage the ease of access to users through the Internet to deploy field experiments. 
These are often performed as A/B tests where two different user groups receive a different treatment, e.g., the old versus new version of a feature.
The treatments are usually assigned randomly to the target groups and statistical analysis is used to determine which of the tested implementations produced better results in terms of improved metrics.
OCEs can be arranged into an ongoing process to implement CE.

\subsection{Implementing Continuous Experimentation}

CE can be seen as arising from advances in the use of experimentation in product development~\cite{thomke1998}, agile software development practices, and continuous software engineering~\cite{fitzgerald2017} practices such as continuous integration, continuous delivery, and continuous deployment.
Modern software engineering practices like these are behind the experimentation processes in pioneering companies such as Microsoft~\cite{kohavi2007,kohavi2009}, Netflix~\cite{amatriain2013,gomez-uribe2016}, and others~\cite{feitelson2013,steiber2013,wu2020}.
Applying experimentation to product development reportedly gives increased customer satisfaction, improved and quantified business goals, and a transformation to a continuous development process~\cite{bosch2012}. Holmström Olsson et al.~\cite{holmstromolsson2012} position experimentation as the last step of an organisational evolution ladder, where agile software development, continuous integration, delivery, and deployment are prerequisites.

Realising CE requires arranging experiments across time.
A number of models have been proposed to prescribe how CE can be implemented as a method.
For example, the HYPEX model describes a detailed experimentation process model spanning the complete product development cycle~\cite{holmstromolsson2014}.
A minimum viable feature (MVF) is implemented and deployed, usage data is collected, and gap analysis is applied to determine the next course of action. 
Experimental results are fed back into the business strategy function, completing the feedback loop.

The RIGHT model describes an experimentation framework in detail~\cite{fagerholm2017}. 
It features a continuous experimentation cycle encompassing business and product vision, strategy, experiment design, and software construction activities. 
It also describes an architectural model for the experimentation infrastructure and related stakeholder roles. 
Furthermore, it covers experiment objects in the form of minimum viable products (MVPs) and MVFs, which are used to rapidly test product hypotheses. 
The RIGHT model assumes that CE capabilities can be built little by little.
Once multiple simultaneous experiments are running, the full model becomes critical.

While other models are available, they share most of their essential characteristics with HYPEX and RIGHT, and the latter is considered a reference model in the literature~\cite{auer2021}.

\subsection{Software Startups and Experimentation}

Software startups are often characterised as being innovative and driven by small teams, operating under high uncertainty and rapidly evolving, and being challenged by a lack of resources, time pressure to deliver a product to market, and only little working and operating history~\cite{paternoster2014,berg2018,melegati2021}. Startups are thus distinctly different from traditional enterprises, in which new product development is only one concern alongside concerns related to existing products and customers.

Startups commonly fail~\cite{paternoster2014}, often by not achieving product-market fit~\cite{ries2011,kotashev2022} -- they create the wrong product~\cite{klotins2021}. To increase the chances of success, the Lean Startup method proposes Build-Measure-Learn loops: testing hypotheses about how users will react to software changes based on predefined metrics, and using the results to make an MVP~\cite{ries2011}. 
Many ideas from the Lean Startup model can be found in continuous software engineering \cite{fitzgerald2017} and CE \cite{holmstromolsson2014,fagerholm2017}.
Camuffo et al. \cite{camuffo2020} conducted a randomised study and concluded that the scientific experimentation approach is beneficial for startups.

Research indicates that while experimentation in general is common in startups, systematic experimentation is not widely adopted~\cite{lindgren2016,melegati2019}.
Previous studies addressing CE adoption (e.g.,~\cite{yaman2017,yaman2016,lindgren2016,holmstromolsson2012}) acknowledge that it requires high levels of skill and coordination, and advanced technological and organisational capabilities.
Auer et al.~\cite{auer2021} identified six categories of challenges:
\begin{enumerate*}[label=\arabic*)]
\item cultural, organisational and managerial challenges,
\item business challenges,
\item technical challenges,
\item statistical challenges,
\item ethical challenges, and
\item domain specific challenges.
\end{enumerate*}
However, few studies address CE specifically for software startups. Although some CE models allow for gradual adoption (e.g., the RIGHT model~\cite{fagerholm2017}), knowledge about systematic adoption strategies and application of the practice in software startups is scarce. Research on CE often describes the benefits, but most studies rely on examples from a small number of industry leaders whose methods are not necessarily suitable for, e.g., smaller companies~\cite{schermann2018} or companies whose products do not share the same characteristics of flexibility. Open questions thus revolve around the circumstances under which CE is suitable for startups and what characteristics of the product development and user data practices of startups are compatible with CE.

\section{Research Method}

We conducted a descriptive multiple-case study~\cite{yin2014} to investigate product development practices in our case companies. Through semi-structured interviews~\cite{merriam2015}, we obtained insights into the product development practices, utilisation of user data, and experimentation used in the companies, which we analysed to develop propositions that address the research questions. We used interviews because they provide relatively quick access to key information, and because other data sources can be scarce in startup companies; for example, extensive process documentation may not exist. Confidentiality requirements prevent us from publishing the full set of collected data, but excerpts are given in this paper. An online supplement gives details on the study design and results~\cite{supplementary-material}.

\subsection{Case Companies}

We selected case companies (see Table~\ref{tab:casecompanies}) from two startup communities in the capital region of Finland:
\emph{Maria 01} (\url{https://maria.io/}), with over 175 startups residing in Kamppi, Helsinki; and
\emph{A Grid} (\url{https://agrid.fi/}), with around 150 startups located on the Aalto University campus in Otaniemi, Espoo.

For the company selection we used purposeful sampling: a convenience sample guided by our research goals~\cite{merriam2015}.
We sought to include companies at any stage of the early product development life-cycle and with different kinds of software-based products or services. Having a relaxed set of selection criteria provided a heterogeneous set of cases and varied qualitative data. All companies remain anonymous and we omit identifying information, such as product details.

\begin{table*}
\caption{Case companies and their context at the time of the study.}
\begin{center}
\begin{scriptsize}
\begin{tabular}{%
    l%
    l%
    >{\raggedright\arraybackslash}p{2.1cm}%
    >{\raggedright\arraybackslash}p{1.9cm}%
    >{\raggedright\arraybackslash}p{3cm}%
    >{\raggedright\arraybackslash}p{3cm}%
}
\toprule
ID & Location & Business model & Distribution channel & Product stage & Participant role \\
\midrule
A & Maria 01 & Free app, sell ads & Mobile app & V1.0 launch in progress & Product owner (PO) \\
B & Maria 01 & B2B (provision\textsuperscript{a}) & API as a service & Released & COO \\
C & Maria 01 & B2C (subscription) & Side-loaded application & Early access program live & CTO \\
D & A Grid & B2B (provision\textsuperscript{a}) & SaaS & Live for over a year & CEO \\
E & A Grid & B2C (subscription) & Physical & Live for over a year & Lead software engineer \\
\bottomrule
\multicolumn{6}{p{0.96\columnwidth}}{\textsuperscript{a}Provision business model: the company receives a provision of sales from their B2B customer.}\\
\end{tabular}
\label{tab:casecompanies}
\end{scriptsize}
\end{center}
\end{table*}
\vspace{-1.3cm}

\subsection{Data Collection}

Semi-structured interviews are suitable when the topic and expected data are not well understood before starting the data collection~\cite{merriam2015}. We expected startups to operate in varied ways that we could not fully anticipate. We designed a semi-structured interview format with open-ended questions based on the guide used by Lindgren \& Münch~\cite{lindgren2015}.
We extended the question set with new questions on tool and practice selection and added CE as a distinct topic.

As shown in Table~\ref{tab:casecompanies}, we interviewed one expert participant from each case company, identified through discussion with company representatives, and based on their broad understanding of both the technical and business aspects of their company. In addition to the main interviews, follow-up data collection with the same or additional participants was conducted to verify and fill in some details. The interviews were performed in Finnish.

\subsection{Data Analysis}

The first author iteratively analysed the interview transcripts. Insights obtained from one case company often provided a new perspective into data from other companies. Multiple passes allowed building higher-order categories from topics that appeared across case companies. The analysis was supported by data review sessions with the two other authors. In these, each transcript was walked through and discussed, with the first author explaining the interview and preliminary insights obtained. Gradually, the sessions moved to discussions about transcript coding and emerging insights from the data. ATLAS.ti was used for the analysis task. Coding was performed in English although the transcripts were in Finnish.

\subsubsection{First-cycle Coding.} The analysis started without any predefined categories. Codes produced from the first analysis iterations were long and descriptive, and multiple coding passes were used to group similar codes under a single parent code. Higher-level categories were built based on recurring themes.

\subsubsection{Individual Case Analysis.} After coding and category-building, each company was analysed as an individual case. Practices for software and product development and user data collection and use were extracted for closer inspection. The information flow within each company was represented graphically, giving a picture of the organisation and the interactions between different stakeholders. 
The graphics are provided in the supplementary material~\cite{supplementary-material}.

\subsubsection{Cross-case Analysis and Second-cycle Coding.} Combined data from all five cases were also analysed. Tools and practices used by the companies were analysed in tabular form. The coded interview data from all five individual cases were re-coded into common themes that permeate the whole data set. In this final phase, the analysis focused on uncovering categories that could potentially explain the tool and practice choices made by the companies.

\section{Results}

Systematic experimentation was rare in the case companies. Our results question the attractiveness of CE for startups as multiple factors detract from its use.

\subsection{Case Company Descriptions}

\subsubsection{Company A.}
Lack of previous experience in mobile application development was a major bottleneck for data acquisition in Company A.
Challenges involved understanding what data was possible to get, what data was valuable, and which tools were worth using.
The application was instrumented to log some user actions, and the team was actively seeking to improve their data collection.

A feedback form was used to collect user data, and app store reviews were a regular feedback channel, with in-app prompts to request reviews. The product owner (PO) reviewed the collected data weekly, but would have wanted to better understand the real-world usage of the application, and have the capability to chat with individual users to gain insights on how and why they used the app.

The PO was not familiar with CE, but mentioned iterative user testing -- resembling experimentation -- for the application’s on-boarding experience.
Think-aloud user tests had been used with a small number of new users. Based on this, design adjustments were made and subsequent user testing indicated an improvement.
The PO considered most of their development as some form of test.
\begin{quote}
Well, almost everything we do is maybe more like tests. (Product owner, Company A)
\end{quote}
However, it is not clear if these tests would qualify as planned experiments,
and whether hypotheses and measurable goals were set before starting development.


\subsubsection{Company B.}
There was only restricted access to user data in this company, as their service was used through a third-party UI.
The company had to rely on the limited data available from their own systems, as the business partners owning the UI were not willing to share detailed user data. 
However, Company B closely collaborated with current and potential customers for requirements elicitation. 
Data from weekly customer meetings were used extensively from early on, and the products were engineered to meet partner-defined functional and non-functional requirements.
Overall, the company appeared to base their product development decisions on the data received through their business partners and less on the scarcely available user data.

Although Company B would likely be technologically capable of running continuous experimentation, the company’s B2B business model makes any experimentation involving user data very difficult.
The company claimed that the pricing of their product was the only parameter they could adjust.

\begin{quote}
We do not have resources at the moment to do anything repeatedly or as a process [\ldots] this interest rate is the only parameter we could control. (COO, Company B, on A/B testing)
\end{quote}

Adjusting the pricing was also considered to be difficult, since any changes would need to be coordinated with, and approved by, their business partners.
These obstacles to experimentation for companies operating in the B2B domain are well recognised in earlier studies~\cite{yaman2017,auer2021}.


\subsubsection{Company C.}
This company was in a transition phase where they had just launched their application through an early access program (EAP).
They used a form of repeated experimentation when building the EAP version, working closely with potential customers and industry experts to understand the requirements their future customers would have for the product. 
After establishing the initial set of requirements, they executed numerous Build-Measure-Learn iterations where each new version of the product was subjected to a user test by an expert user, including observation and an interview for feedback.
The collected user data was then converted into software development tasks. 

\begin{quote}
It has been this [\ldots] very tight develop and test type of work where we develop something and then go [try it out]. (CTO, Company C, on the development process)
\end{quote}

However, there was no sign of predefined metrics or evaluation criteria which would have indicated a systematic approach for experimentation. The company had thought about possible future A/B testing and had recently implemented feature flags, but mainly for licensing purposes.
Instrumentation data was not automatically sent to the company due to user privacy expectations and offline capability requirements.
The company expected that most feedback would be received via email and discussing with the users.
This was seen as good enough for now and they recognised that some other way of handling user feedback would need to be implemented once their customer base grows.


\subsubsection{Company D.}

This company used a well-defined process for product development planning, following the Data-Insight-Belief-Bet (DIBB) framework~\cite{kniberg2016}.
The leadership defined a set of objective key results (OKR) for the next quarter, to be implemented by the product development teams.
Each OKR contained a high-level goal and the metrics to evaluate whether or not the goal was reached at the end of the quarter. The focus was on meeting investor goals.

The development teams regularly built feature prototypes, mostly used internally, but occasionally presented to end users or customer representatives.
An expert board reviewed the data from prototyping and selected the best implementation based on qualitative feedback.
The data was used continuously as company D was not only reacting to negative feedback, but also actively trying to learn from the data and even predicting future needs based on historical data.

\begin{quote}
Ten percent of the features [have come from when we analysed all feedback] and predicted what kind of things there would be coming. [We can say] we actually need to build this kind of feature because [customers] are going towards this kind of world. (CEO, Company D, on use of data)
\end{quote}

The company employed numerous user data collection channels and used data extensively in all stages of product development.
Event-level quantitative data was collected from different parts of the system and qualitative data was collected through a chat embedded in the service.
Research ops, a dedicated business unit, analysed and refined all available data, and was responsible for the data to be available to all members of the organisation.


\subsubsection{Company E.}
The last company offered a physical service; their software provided internal support.
There was no instrumentation for quantitative data collection due to lack of interest in such feedback.
To collect qualitative user data, the company mainly observed office staff and read feedback from chat channels.
This internal feedback loop tied back to the development team's Kanban board where new input was entered as work items.
A survey was used to collect feedback from company field workers, but it was not entirely clear how the data was used, suggesting a lack of systematic usage.

\begin{quote}
We have sent our field workers these quick surveys just to see what is the general opinion about this app and such. We did not have any specific metrics like ``rate this from 0 to 10'' or anything. (Lead developer, Company E, on user surveys)
\end{quote}

Feedback on the physical service was collected through a form and email, was handled by customer service agents, and was generally not available to everyone in the company.
Overall, Company E did not systematically use data in the internal software product development.
We found indications that data was seen as a way to detect problems rather than understanding users' needs.

\begin{quote}
[\ldots] often not receiving any feedback about anything indicates that things are probably quite ok as there is nothing to complain about. (Lead developer, Company E, on negative feedback)
\end{quote}

The team had envisioned doing A/B testing once they have released customer-facing applications.
However, in its current state, the company lacked the capabilities to effectively use experimentation since the software development team was largely disconnected from the rest of the organisation.


\subsection{Cross-case Analysis}
\label{sec:cross-case-analysis}

All case companies had an incremental software development process and used lean and agile practices such as continuous integration and delivery. They also had several prerequisites for continuous deployment in place but none of them actually deployed automatically to production.

Different forms of experimentation were used in four out of five companies.
Company A did user testing with different UI implementations and a MVP feature implementation. A price change experiment was going on in Company B, prototyping and expert user testing in a continuous cycle in Company C, and extensive prototyping and user testing in Company D. Company C appeared to be using systematic prototyping, but their data collection and data analysis practices were not systematic: they simply let expert test users try the product without planning what data to collect. Company D had advanced data collection, storage, and refinement capabilities, used the data systematically, and would most likely be capable of continuous experimentation.

It appeared that the startups had set up their tools and practices based on previous experience of founders or employees.
There was isolated use of practices such as user testing, prototyping, and even occasional A/B testing in the companies, but those seemed to originate from individuals with previous experiences with the techniques rather than the company actively adopting the practices.

Allocating resources to execute the product roadmap in the most efficient way was central to work planning in all case companies. Only Company B did not mention funding as a resource limitation. All companies had roadmaps with far more work than they could ever implement given their available schedule and resources. This chronic lack of resources forced the companies to prioritise work that would be beneficial in the next funding round, even at the expense of feature development with real value to customers.

There was a tendency in the companies to avoid advanced practices until a strong need emerged -- a fiercely pragmatic stance towards method adoption. It became apparent that even experiencing problems was not enough to adopt better ways to operate. Even if the team would know a better way of doing their work, the effort to set up a new system and the opportunity cost of being unable to work on other, more important, tasks, could prevent change from taking place. Change was seen as a waste of resources unless a perceived tangible benefit was in sight within the time horizon that is currently relevant for the company.

\section{Discussion}\label{sec:discussion}

The descriptive account given above allows us to construct a number of propositions to address the research questions, which we now discuss.

\subsection{CE and Software Development Methods (RQ1)}

To address RQ1, we propose that \textit{software startups do not generally integrate experimentation into their software development methods, but may use CE-support{-}ing methods}. Experimentation in a loose sense was present in our case companies, but systematic, statistically controlled experimentation was uncommon, in line with existing research showing a lack of systematic CE in startups~\cite{lindgren2015,gutbrod2017,melegati2019}. Product domains and business models that limit the possibilities for experimentation further reduce integration. Maturing startups may introduce more structure into their methods, and we then expect systematic experimentation to be more frequent and deeply integrated into methods and practices.

Several practices supporting CE were in place in the case companies. All developed their products incrementally using agile or lean software development practices. Only Company A built their mobile app manually. The companies were similar in development practices, the single major difference being automated testing, used by only two.
Version control and backlog were the only practices with an adoption rate above 35\% in a prior study~\cite{klotins2021}.
Compared to this, the case companies are fairly advanced in adopting key agile practices. However, a closer look reveals rather selective adoption of agile practices in three of our case companies, and some deviations were found in all. Method selectiveness has been reported before, with a warning that picking only some agile practices without supporting practices may lead to adverse effects~\cite{klotins2021}.

The prerequisites for experimentation thus appear at first glance to be in place in the case companies, but they have picked methods selectively. Experimentation does not appear to be integrated with the methods.
We found that only one company had a systematic approach to experimentation, which is in line with findings from a study with ten case companies \cite{lindgren2015} reporting that non-systematic experimentation was common, but systematic experimentation among startup companies was rare.

\subsection{Use of Data (RQ2)}

For RQ2, we propose that
\begin{enumerate*}[label=\arabic*)]
\item \textit{utilisation of user data in startups tends to begin with ad hoc collection and interpretation of rich, qualitative data when the user base is small and the startup is in its early stages,} but that
\item \textit{transitioning to more systematic use of data, which allows reliable understanding of the effects of product decisions, requires deliberately building specific technical capabilities and adopting an experimentation framework.}
\end{enumerate*}

Four of the five companies had good access to user data and established ways for collecting qualitative feedback. Interviews, feedback forms, and email were the most common channels, but social media, application store feedback, and chat were also used. Company B's B2B model practically prevented access to qualitative end user data, and forced them to rely on data their business partners were willing to share. These are known issues of the B2B context \cite{rissanen2015}.

Four of the companies had built instrumentation for collecting event-level quantitative data, but only companies A and D had automated access to UI level events; both demonstrated product improvement as a result of using this data.
Company B was limited to backend data, and Company C had implemented an opt-in data collection mechanism primarily for debugging purposes.

Prior work suggests that CE adoption relies on initial awareness~\cite{gutbrod2017}, and progresses through several stages, each with many challenges to overcome~\cite{melegati2022}. However, we add the consideration that CE in the form often proposed in the literature may be inadequate for startups as its costs may be too large.
Our findings show that, in line with previous research on organisational development towards CE (e.g., \cite{bosch2012,holmstromolsson2012,fitzgerald2017}), the transition to systematic and advanced use of data in startups requires considerable investment in development of skills, procedures, data acquisition, and customer collaboration in different forms. 
It requires a mode of functioning that is not easily combined with the sparse resources of startups nor with the culture of fast-moving innovation and a sharp focus on realising the founders' product vision.
Adopting a systematic experimentation approach may face resistance (c.f.~\cite{melegati2022}), pointing to the need for data acquisition and utilisation methods that work for very early-stage startups, and can scale up as they grow and mature.

\subsection{The Appeal of Continuous Experimentation (RQ3)}

The following three propositions concern conditions for adopting CE in startups.

\noindent \emph{P1. Adoption of a practice requires previous personal experience.}
Only the interviewee from Company D had heard about continuous experimentation, but was unable to describe it in detail, which suggests that they had no prior experience with it. Our findings suggest that the strongest adoption pathway in this context is word of mouth and personal experience.

\noindent \emph{P2. The practice must not require large resources to adopt or use.}
Four out of the five companies mentioned funding as critical to their their priorities.
The fear of running out of time and money may prevent startups from taking risks in adopting practices, especially if large time and resource investments are perceived to be needed.
It is unclear what resources are required to adopt CE.
The case companies had already performed experiments, albeit in unstructured ways.
Becoming more systematic could be a matter of educating the company employees. 
However, establishing an experimentation process and the required technical infrastructure requires upfront work.
Additionally, more resources are needed to make variants for experiments than to build a single version. 
These resource requirements may exceed what the companies believe they can afford.

\noindent \emph{P3. The practice must solve a serious enough problem within a reasonable amount of time.} 
Whether CE solves a concrete enough problem and provides enough perceived value is a question beyond personal experience and resource requirements.
From the company's perspective, an issue must be serious enough and timely, and the new practice must guide to a solution fast enough to keep up with the speed dictated by investors.
At least companies A and C, which had not yet established a firm user base, were effectively looking for a product-market fit and trying to test their ideas with early versions of their application.
These two companies seemed to be less inclined to do systematic experimentation than they were to develop features from the roadmap and see what users would say about the next version. 
Even though CE could have some benefits, these companies may be more interested in increasing the speed of development.
Therefore, the companies would not recognise CE as a valuable solution for their problem.

Earlier studies covering CE adoption (e.g.,~\cite{holmstromolsson2012,lindgren2016,fagerholm2017,gutbrod2017,yaman2019,melegati2019,melegati2022}) indicate that notable skill and coordination from the whole organisation is required.
They indicate that adopting CE is a journey that should be taken gradually, starting with small-scale experimentation, and building increased technical and organisational capabilities in each subsequent experimentation round~\cite{lindgren2016,yaman2019}.
Yaman et al.~\cite{yaman2019} propose having a CE champion with the required expertise and mandate to facilitate adoption. 
This idea aligns well with the identified need for prior experience, as well as the startup companies' suggested inability to see the value of systematic experimentation. 

\subsection{Implications for Practice}

The propositions above suggest that CE is problematic for software startups. On one hand, the possible benefits, especially the validation of value hypotheses, are important for startups since they help avoid directions that do not lead to a viable product. On the other hand, the considerable resources required for full adoption, the many details and required rigour in currently existing methods, and the potential mismatch with fast-paced startup culture and the skills that startup employees currently have, mean that adopting CE risks depleting precisely those scarce resources that startups need to build their product.

As noted above, the current solution is to do a gradual, piecemeal adoption, and to find a person or small team to spearhead the practice, using minimal resources. However, we suggest that startups should consider CE without all the prerequisites that are usually listed. Simple means of observing user behaviour, such as sending back a single value from an app, and basic analyses in a spreadsheet, can enable startups to focus on product questions rather than the infrastructure. This turns the order of adoption around to provide immediate value: instead of large up-front investments in CI, CD, and other automation that must be constantly updated as the product changes, the focus should be on developing the capability to dress product development questions as simple experimental designs and finding the quickest and least resource-intensive ways to execute them. This leverages the key assets that startups have: knowledge about their product and customers, a sense of the product vision, and an aptitude for fast-paced innovation, turned towards the practice of experimentation itself.

\subsection{Limitations}

The credibility or internal validity~\cite{merriam2015} of this study is limited by the participation of only one main interviewee per company. To counter this threat, we sought to include participants who had the best possible knowledge of the questions we wanted to ask. We also included a small number of additional participants to provide missing details on, e.g., technical matters. Credibility, as well as consistency~\cite{merriam2015} is also strengthened by our use of researcher triangulation and the repeated data sessions carried out during the analysis. Given the small size of the companies, we argue that our study is credible and consistent in capturing the reality of each of the five companies.

In terms of transferability or external validity~\cite{merriam2015}, the variation in case companies and the reporting of results and circumstances should help the reader determine the extent to which our findings can be applied in specific future cases. We have not sought to statistically validate the findings in a larger sample. Rather, the aim of this study is to develop a descriptive account of the case companies and develop propositions that could be used, for example, to design a larger survey study to examine the prevalence of the factors found here. We argue that the propositions obtained in this study enable future studies to examine CE in startups in more detail than what was possible based on existing studies to date, and that they can be used as points of reflection for practitioners if they are considering to adopt CE in a startup context.

\section{Conclusions}

We sought to understand the factors involved in adopting CE in software startups. Through a descriptive multiple-case study in five Finnish startups, we examined product development practices, and method choices, asking how these companies collected and utilised user data in product development.

All companies used agile or lean software development practices and continuous software development practices such as continuous integration and continuous deployment.
Most companies were able to collect qualitative user data. Two were also doing automatic collection of user interface event level data from their services. The companies used the collected data mostly for validating that their services did not contain errors. Only one company appeared to be systematically using the data for predicting future user requirements.

Continuous experimentation was not commonly known: only one participant had heard about it and could not describe it in detail.
Previous experience and expected short-term value seem to be important factors when startups select CE tools and practices.
The companies struggled with limited resources, forcing them to carefully prioritise work and foregoing the adoption of complex methods.

A well-resourced company can afford dedicating an extra team to experimentation without negatively affecting the development work. This is not feasible for a startup with only a handful of developers, where the effort of doing an experiment might require halting other development. Thus, it is understandable that a startup company would prefer to simply continue executing their roadmap.

More awareness of continuous experimentation could improve the adoption rate of the practice among developers and entrepreneurs. Potential approaches include teaching the practice in university curricula or to prepare entrepreneurship training programs with CE-related material. The latter approach has been tried with promising results~\cite{camuffo2020} and may be worth pursuing in further research.

However, the CE practice should also adapt to the requirements of different kinds of companies. The research community should seek ways to make CE more affordable for a larger variety of companies, to lower the adoption barrier on both organisational and individual levels, and to make it more attractive and easier to benefit from the practice's advantages.

\subsubsection{Acknowledgements}
We express our gratitude to the study participants and participating companies for their generous sharing of information.

%
%
%
\bibliographystyle{splncs04}
\bibliography{citations}
\end{document}